\documentclass[amsmath,amssymb,aps,floatfix,prd,a4paper]{revtex4}

\usepackage{dcolumn}
\usepackage{bm}
\usepackage{epsfig}
\usepackage{amsfonts}
\usepackage{amssymb,amscd}
\usepackage{array}

\usepackage{graphicx}
\usepackage{caption}

\def\lsim{\raise0.3ex\hbox{$<$\kern-0.75em\raise-1.1ex\hbox{$\sim$}}}

\def\gsim{\raise0.3ex\hbox{$>$\kern-0.75em\raise-1.1ex\hbox{$\sim$}}}

\newcommand{\be}{\begin{equation}}

\newcommand{\ee}{\end{equation}}

\def\beq{\begin{equation}}

\def\eeq{\end{equation}}

\def\beqa{\begin{eqnarray}}

\def\eeqa{\end{eqnarray}}

\def\la{\langle}

\def\ra{\rangle}

\newcommand{\ba}{\begin{eqnarray}}

\def\gappeq{\mathrel{\
rlap {\raise.5ex\hbox{$>$}}

{\lower.5ex\hbox{$\sim$}}}}

\def\lappeq{\mathrel{\rlap{\raise.5ex\hbox{$<$}}

{\lower.5ex\hbox{$\sim$}}}}

\def\Toprel#1\over#2{\mathrel{\mathop{#2}\limits^{#1}}}

\begin{document}

%\begin{flushright}
%LU TP 15-XX\\
%December 2015
%\vskip1cm
%\end{flushright}

\title{Multiplicity distributions in the low x regime in a simple model} 
%\author{G. R. Germano  and  F.S. Navarra}
\author{G. R. Germano}    
\email{guilherme.germano@usp.br}
\affiliation{Instituto de F\'{\i}sica, Universidade de S\~{a}o Paulo,
  Rua do Mat\~ao, 1371, CEP 05508-090 , Cidade Universit\'aria,
  S\~{a}o Paulo, SP, Brazil}

\author{F.S. Navarra}
\email{navarra@if.usp.br}
\affiliation{Instituto de F\'{\i}sica, Universidade de S\~{a}o Paulo,
  Rua do Mat\~ao, 1371, CEP 05508-090 , Cidade Universit\'aria,
  S\~{a}o Paulo, SP, Brazil}

\begin{abstract}
  
In this work we introduce small changes in the model proposed by E. Levin
and D. Kharzeev for multiplicity distributions of  particles produced in
proton-proton collisions. We compare the predictions of the model  
with the available experimental data from the LHC. We also consider the most
recent version of the model proposed by E. Gotsman and E. Levin. These two 
versions of the model give a good description of the measured multiplicity
distributions in the  central pseudo-rapidity region ($|\eta| < 0.5$).
The agreement with data is not so good when we consider the particles 
measured over a wider pseudo-rapidity interval ($|\eta| < 2.4$).
Interestingly, the agreement becomes
better  when one selects  events with higher transverse momentum
($p_T  >  500$ MeV). 

\end{abstract}

%\pacs{12.38.-t, 24.85.+p, 25.30.-c}

%\keywords{Quantum Chromodynamics, Double Vector Meson Production,
%Saturation effects.}

\maketitle

%\vspace{1cm}

\section{Introduction}

The most recent review paper on multiplicity distributions  was written
ten years ago \cite{fiete}. Its last section presented a number of
predictions for the future collisions that would be performed at the LHC. 

After ten years of operation, the
LHC delivered an impressive amount of data on multiplicity distributions
\cite{atlas11,cms11,atlas16a,atlas16b,atlas16c,alice17,cms18}.  The collision
energies range from 0.9 to 13 TeV. These data allow us to study the energy
evolution of multiplicity distributions and  to look for qualitative changes 
in the reaction dynamics.  One of the expected changes was a transition from
the soft (non-perturbative) to the semi-hard (perturbative) QCD regime 
\cite{ugo98}. Previous studies had concluded that this transition was     
already happening \cite{fiete} in the pre-LHC era. It was conjectured that 
semi-hard events would have larger multiplicities and this would
induce  a ``shoulder'' in the large $n$ region of the multiplicity distribution
$P(n)$. Up to $\sqrt{s} = 1$ TeV, $P(n)$ was well described by a negative
binomial distribution (NBD). At the Tevatron at $\sqrt{s} = 1.8$ TeV the
appearance of the ``shoulder'' was confirmed and this led to the two-component
model proposed in \cite{ugo98} in which the measured data were described by a
combination of two NBDs,  one representing the  soft and one the semi-hard
component. This ``shoulder'' was also responsible for the  violation of the
Koba-Nielsen-Olsen (KNO) scaling \cite{fiete}.  More recently, the data taken
at $\sqrt{s} = $ 7, 8 and 13 TeV show that the best fit is obtained with the
inclusion of a third negative binomial distribution \cite{zbo,bya}.

At the LHC another transition was expected to be seen:  the transition to the
low x regime and to the BFKL dynamics with the possible manifestation of gluon 
saturation effects \cite{kole,gel}.  
Compared to the previous accelerators, the LHC proton-proton collisions occur
at a center of mass energy which is one order of magnitude higher. Increasing
the energy we have access to the low momentum component of the proton wave
function, i.e., we ``see'' more partons with smaller momentum fraction $x$
of the
proton. At the same time, we may have collisions between partons which are more
off-shell, i.e., which have a higher virtuality $Q^2$. The changes in the
parton
distribution functions $f(x,Q^2)$ when they enter in the low $x$ and high $Q^2$
regimes are described by the BFKL and DGLAP QCD evolution equations
respectively. The solutions of these equations show that the number of partons
increases with $1/x$ and with $Q^2$. We can picture this process thinking
that when
we boost a proton its partons go through a branching process, with a copious
production of new partons. This has direct implications for the final particle
multiplicity. At lower collision energies particle production is dominated by
string fragmentation, which describes hadron formation and which
is a non-perturbative QCD process. At higher energies, and especially at the
LHC, particle production is dominated by parton branching, which is studied
with perturbative QCD, with string fragmentation playing a secondary role. The
measured multiplicity distributions at the LHC are thus an excellent testing
ground for the predictions of the evolution equations. 

The best theory of strong interactions is QCD.
The part of QCD which describes dense partonic systems formed in high energy
collisions is the Color Glass Condensate (CGC) theory \cite{kole,gel}.
In practical applications of the CGC  it is necessary to make
approximations and/or additional assumptions to obtain results which can be
compared to data.  In several works it has been shown that with the
CGC based  models it is possible to obtain a reasonable description of the
global features of multiparticle production in high energy collisions.
In the case of multiplicity distributions there is a lot of work to be done.
In \cite{schenke} the authors used the IP-Glasma model and calculated the multiplicity
distribution of particles produced with $|\eta|<0.5$ in pp collisions at
7 TeV. They found discrepancies between the model predictions and data at
large multiplicities. 
In \cite{nara,raju11} the authors used CGC models to study particle production 
and $P(n)$  was assumed to be a negative binomial distribution with the
parameters $k$ and $\bar{n}$ given by the CGC models. The data analysis was
restricted to the lower LHC energies and to central pseudo-rapidity intervals.
In the present analysis we will include the most recent data from higher
energies and consider other rapidity and $p_T$ windows. Moreover we will
test a $P(n)$ which is related to BFKL dynamics.

In \cite{glitter} it was shown that the number of particles produced from
the glasma follows a negative binomial distribution, which depends on the
parameter $k$, which turns out to be proportional to the saturation scale
$k \propto  Q_s^2 $. We know that
the saturation scale grows with the collision energy and hence so does $k$.
As $k \to \infty$ the negative binomial distribution goes to a
Poisson distribution, which is a very narrow distribution. Therefore the
CGC-Glasma prediction is that at increasing energies the multiplicity
distribution should ``shrink''. So far, the existing data have shown the
opposite trend. The observed broadening of the multiplicity distributions
can be quantified through the moments $C_n$  which are increasing functions
of the energy, implying broader multiplicity distributions. The
non-observation of the shrinkage of $P(n)$ is an indication that some
ingredient is still missing in the theory and it must be completed.

In  theoretical studies of MD's there are top-down and bottom-up
approaches.
In the first group we find works \cite{glitter,mumu,giaca,dumi,shab},
firmly rooted in QCD, which discuss in detail the
parton branching process using different formalisms, different techniques and
different approximations. Special attention is given to  gluon saturation,
which takes place at very high gluon densities, when gluon-gluon fusion becomes
important  and reduces the growth of gluon number, having a direct impact on
the final hadron multiplicity. In a non-perturbative approach, it was
suggested in  \cite{khoze} that the presence of instantons could affect the
observed multiplicity distributions. 

In the second group we find works \cite{zbo,bya,beg,agga,sha,gre18,gre19,gre20}
that try to
extract the maximum amount of information from data, using sophisticated
statistical tools,  paying attention to the subtleties
of the measured distributions and building up models, with assumptions that may
motivate an a posteriori theoretical study.

In the top-down approach most of the works did not reach the point where they
address the data and try to reproduce them. In the bottom-up approach most the
works did not reach the point where they establish a clear connection to QCD.

In a third group, we find works in which the authors try to build the bridge
between QCD and data. In this cathegory we include the  studies performed with 
more phenomenological models \cite{nara,raju11,adrian,duna}
and with event generators
\cite{cms11,lund}, which are built to reproduce all the observable quantities
in a high energy collision.

The accumulated experimental data were studied with bottom-up models and with
phenomenological models. The global features of the MD's are captured by the
models but they fail to reproduce the data in one or another kinematical
region. In
the compilation of the theoretical results presented in \cite{cms11}, the event
generators either underestimate or overestimate the large $n$ tail of the
distribution $P(n)$. Inspite of the recent improvements, such as, for example,
in \cite{lund}, the discrepancies persist. There is not yet a complete and
satisfactory theoretical description of multiplicity distributions.

In this work we try to follow the  approach of the third group: we choose a
simple model of the QCD branching process and perform a comprehensive       
comparison with the existing data.  This model was first proposed some time
ago \cite{lubi}, it was
used recently \cite{kale} to study entanglement entropy in particle production
and it was improved very recently in \cite{gole,gole2}, where a preliminary
discussion of the LHC data was presented.  In the next section we briefly
describe the model and in the last section we present our results.

\section{The model}

\subsection{The Kharzeev - Levin model}

In Ref. \cite{lubi,kale}, the authors have developed a model  for multiplicity
distributions based on the BFKL equation, which we will call KL model. 
They propose the following
evolution equation for the parton multiplicity distribution $P_n$:
\beq
\frac{d P_n(Y)}{d Y} \, = \, - \, \Delta \, n \, P_n  \, + \,
(n-1) \, \Delta \, P_{n-1} (Y)
\label{kl}
\eeq
which has the simple solution:
\beq
P_n(Y) \,  = \, P_{KL}(n) \, = \, 
e^{-\Delta Y} \left(1 - e^{- \Delta Y} \right)^{n - 1}
\label{klpn}
\eeq
where $Y = ln(1/x)$ and $\Delta$ is the BFKL Pomeron
intercept, $\Delta\, = \,  4 \,\,ln \,\,  2  \,\,  \bar{\alpha}_s$ with 
$\bar{\alpha}_s = \alpha_s N_c / \pi$. From the above expression we obtain
the mean multiplicity: 
\beq
\langle n  \rangle  = \sum_n n \, P(n) = e^{\Delta Y} =
\left( \frac{1}{x} \right)^{\Delta}
%= \left( \frac{s}{q_0^2} \right)^{\Delta}
\label{nbfkl}
\eeq
The variable $x$ is defined here as in \cite{lelu,babi}:
\beq
x = \frac{q_0^2}{s}
\label{defex}
\eeq
where $q_0$ is a constant. 
Inserting (\ref{defex}) into (\ref{nbfkl}) we obtain:
\beq
\langle n  \rangle  =  \left( \frac{s}{q_0^2} \right)^{\Delta} 
\label{nf}
\eeq
The energy scale $q_0$ can be a mass or the average transverse momentum and
hence it might be different for different data sets, but it should not
depend on the collision energy $\sqrt{s}$. 

The above expression is a prediction of the model. Interestingly, it can be
derived from the CGC formalism.  In fact, 
the first estimate of the mean multiplicity of produced particles in the
CGC framework was done in \cite{kn} and in  \cite{kln}.  According to
\cite{kn} (see also \cite{adrian})
in the saturation regime the number of produced partons is given by
\beq
\langle n \rangle \,   = C \, \frac{Q_s^2 \,  R^2}{\alpha_s(Q^2_s)}
\label{nkn}
\eeq
where $C$ is a constant, $R$ is the transverse size of the projectile
(proton in our case)  and $\alpha_s(Q^2_s)$
is the strong coupling evaluated at $Q^2_s$, which is the saturation scale
given by
\beq
Q_s^2 = Q^2_0 \left( \frac{x_0}{x} \right)^{\lambda}
\label{qs}
\eeq
where $\lambda = 0.2 - 0.3$, $Q_0^2$ and $x_0$ are constants.
From the above equations we conclude that:
\beq
\langle n \rangle  = \, C \, Q_0^2 \, (\frac{x_0}{q_0^2})^{\lambda} \,
\frac{R^2(s)}{\alpha_s(s)} \, s^{\lambda}
\label{nkn2}
\eeq
The radius of the proton, $R$,  is related to confinement
and can not be estimated from CGC physics. It has some weak energy dependence
which comes from nonperturbative effects. In \cite{dosch} the proton
radius was parametrized as
\beq
R^2(s) = R^2_0  \left(\frac{s}{s_p}\right)^{0.05}
\label{prad}
\eeq
Assuming that $\alpha_s(s) \simeq const$ and substituting (\ref{prad})  into
(\ref{nkn2}) we find
\beq
\langle n \rangle \,  = \, \left( \frac{s}{s_0} \right)^{\lambda_{eff}}
\label{nkn4}
\eeq
which is equivalent to (\ref{nf}) and where all the constants were
packed in the parameter $s_0$ and $\lambda_{eff} = \lambda + 0.05$.

At lower energies,
$\sqrt{s} = 10 - 50$ GeV,  particle production is dominated by soft,
non-perturbative dynamics, and  the calculated quantities have typically
a weak (log) energy dependence. In the $\sqrt{s} \simeq 1$ TeV region 
it was argued \cite{ugo98}  that the
mean multiplicity has a soft $\la n_s \ra$ and a semi-hard $\la n_h \ra$
component which depend on the collision energy $\sqrt{s}$ as
$\langle n_s \rangle  \propto ln s $ and $\langle n_h \rangle \propto ln^2 s $.
At the LHC energies it was shown in \cite{cms11} and \cite{kln}  that power
law expressions such as (\ref{nf}) and (\ref{nkn4}) are able to reproduce the
data. However these data are also compatible with other parametrization forms,
such as $\langle n_{ch} \rangle  \propto ln^3 s $, as was shown in \cite{roy}.

In the next section we will compare (\ref{klpn}) with  the available
experimental data from LHC. Eq. (\ref{klpn}) depends on two parameters,
$\Delta$ and $q_0$, which will be adjusted to fit the data. Before doing
this we will describe the improvement of the KL model proposed in
\cite{gole} and \cite{gole2}.

\subsection{The Gotsman - Levin model}

The formalism developed in \cite{kale} was further explored in
\cite{gole}. 
In \cite{gole} it was pointed out that (\ref{klpn}) yields multiplicity
distributions which are too broad. This behavior was attributed to the fact
that (\ref{klpn}) describes dilute systems. According to the authors, a
more accurate description of central rapidity data, requires the study
of the collision between dense systems.
%
% kt fac X hybrid gluons X quarks
% inelasticity 
%dense-dense interactions was performed with the help of the effective
%Hamiltonian studied in a previous paper \cite{kovner} and resulted in
%the following multiplicity distribution: 
In \cite{gole2}, Gotsman and Levin derived the following formula for
the multiplicity distribution:
\beq
P_n(Y) = P_{GL}(n) \, = \, \frac{2}{\sqrt{\pi z}} \frac{e^{-z}}{N}; 
\label{glpn}
\eeq
where $z = \frac{n}{N}$ and
\beq
N = e^{\Delta Y} = e^{\Delta ln (\frac{1}{x})} =
e^{ln(\frac{s}{q_1^2})^{\Delta}}
= \left(\frac{s}{q_{1}^{2}}\right)^{\Delta}
\label{ngl}
\eeq
It is useful to write the simple relation between $P_{KL}(n)$ and
$P_{GL}(n)$ \cite{gole2}:
\beq
R = \frac{P_{GL}(n)}{P_{KL}(n)} = \sqrt{\frac{1}{\pi z}}
= \sqrt{\frac{N}{\pi n}}
\label{ratio}
\eeq
This more singular behavior of $P_{GL}(n)$ at small $n$ will be visible
in all the figures shown below.

\section{Results and discussion}

In this section we compare the results of the model with the experimental data  
from LHC.  Three data sets were used, each one with several  center of mass
energies ($\sqrt{s}$) and each set corresponding to a different cut in the
minimum transverse momentum ($p_{T}$) of the measured particles and a different
pseudo-rapidity ($\eta$) window. They are: 
\begin{itemize}

\item Set  I: $p_{T} > 100 $ MeV, $|\eta|<0.5$ and energies      
$\sqrt{s} =$ 900, 2360 and 7000 GeV. These data are from \cite{cms11}. 

\item Set II: $p_{T} > 100 $ MeV, $|\eta|<2.4$, and energies
$ \sqrt{s} = $  900,  7000 and 8000 GeV \cite{alice17} and  8000
\cite{atlas16a}  and 13000 GeV \cite{atlas16b}.

\item Set III: $p_{T} > 500 $ MeV, $|\eta|<2.4$ and energies 
$\sqrt{s} = $ 900, 2360 and 7000 GeV \cite{atlas11}, 
8000 GeV ($|\eta|<2.5$) \cite{atlas16a}
and 13000 GeV \cite{cms18}
and ($|\eta|<2.5$) \cite{atlas16c}.

\end{itemize}

\begin{figure}
\begin{tabular}{cc}  
  {\includegraphics[width=.5\linewidth]{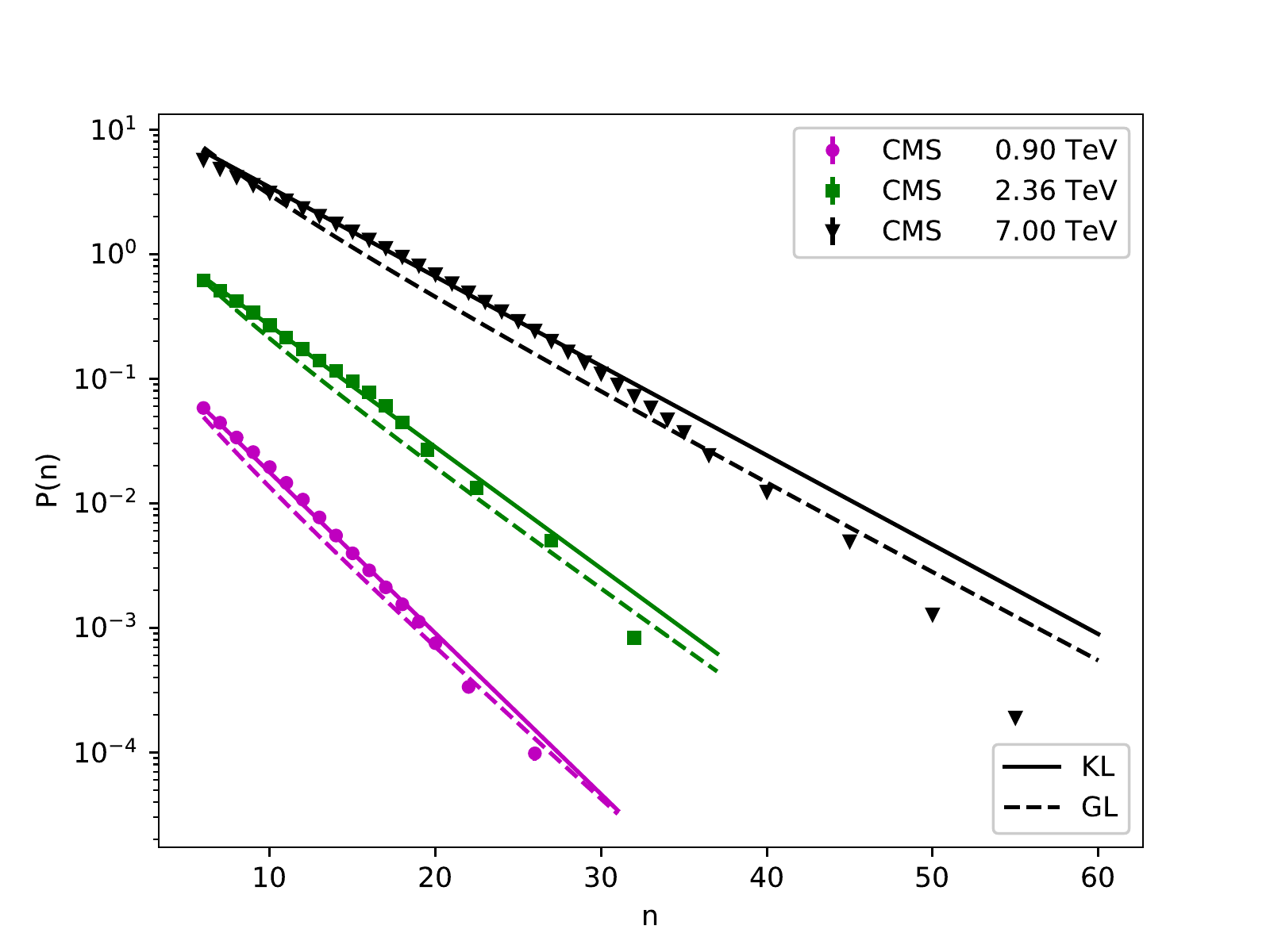}} &
  {\includegraphics[width=.5\linewidth]{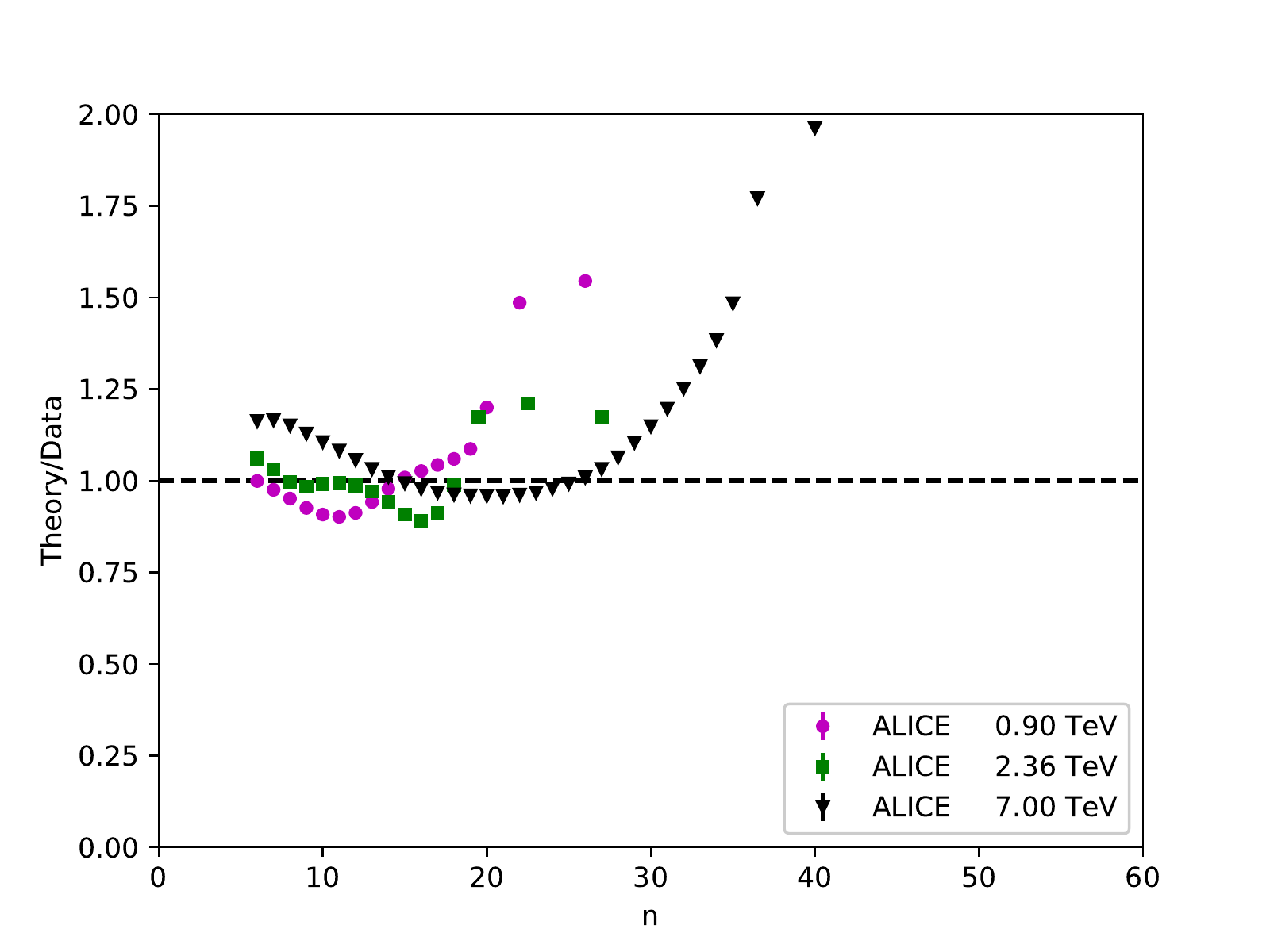}}\\
  \end{tabular} 
\caption{Multiplicity distributions compared with data set I. 
  Data are from CMS (2011) \cite{cms11}.
  Left panel: $P(n)$ for different energies. Solid lines: $P_{KL}(n)$.
  Dashed lines: $P_{GL}(n)$. 
  Sets of points were multiplied by powers of ten for clarity. Right panel:
  Ratio Theory ($P_{KL}(n)$) /Data.}
\label{pnmeio}
\end{figure}

\begin{figure}
\begin{tabular}{cc}
  {\includegraphics[width=.5\linewidth]{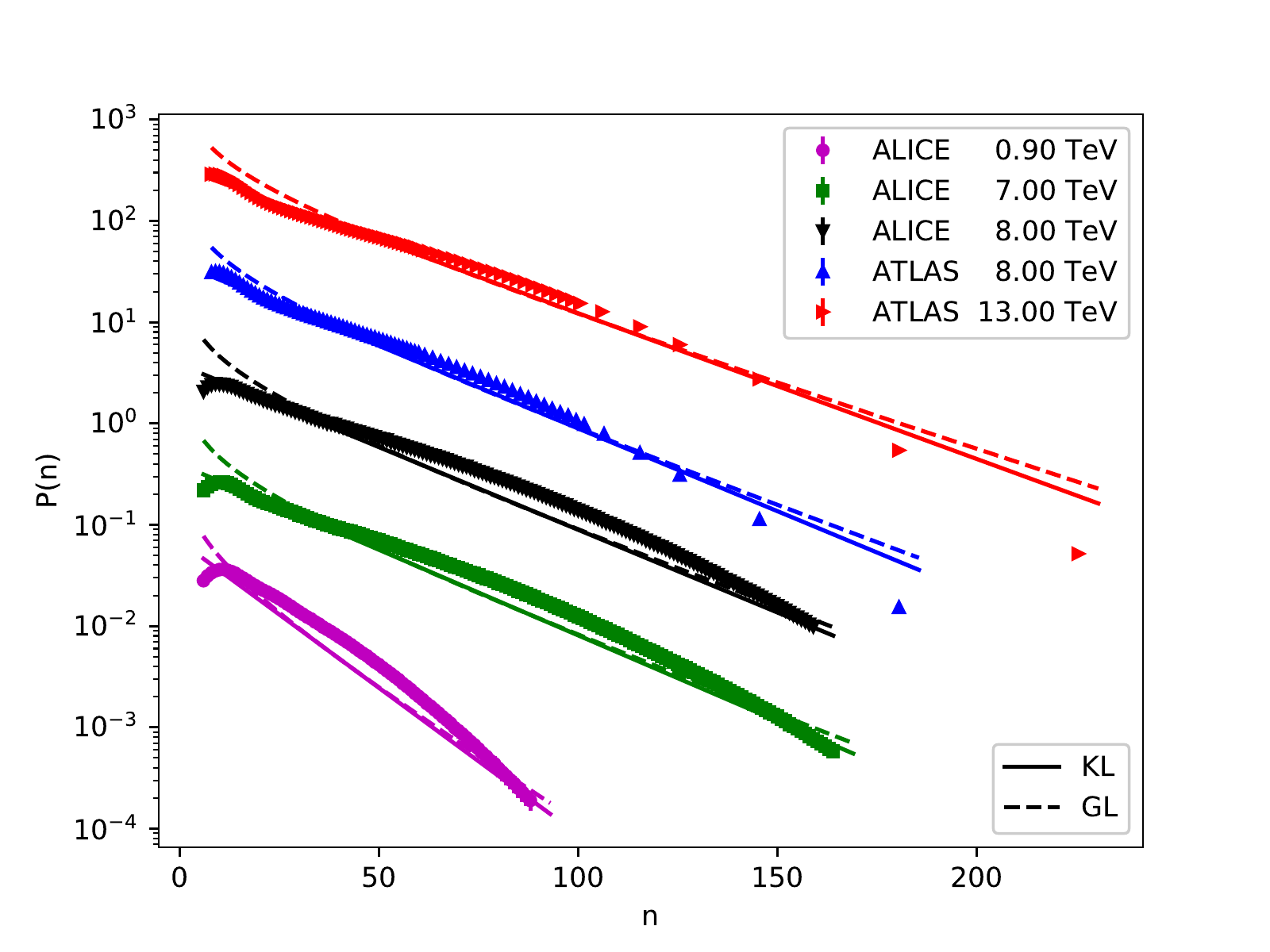}} &
  {\includegraphics[width=.5\linewidth]{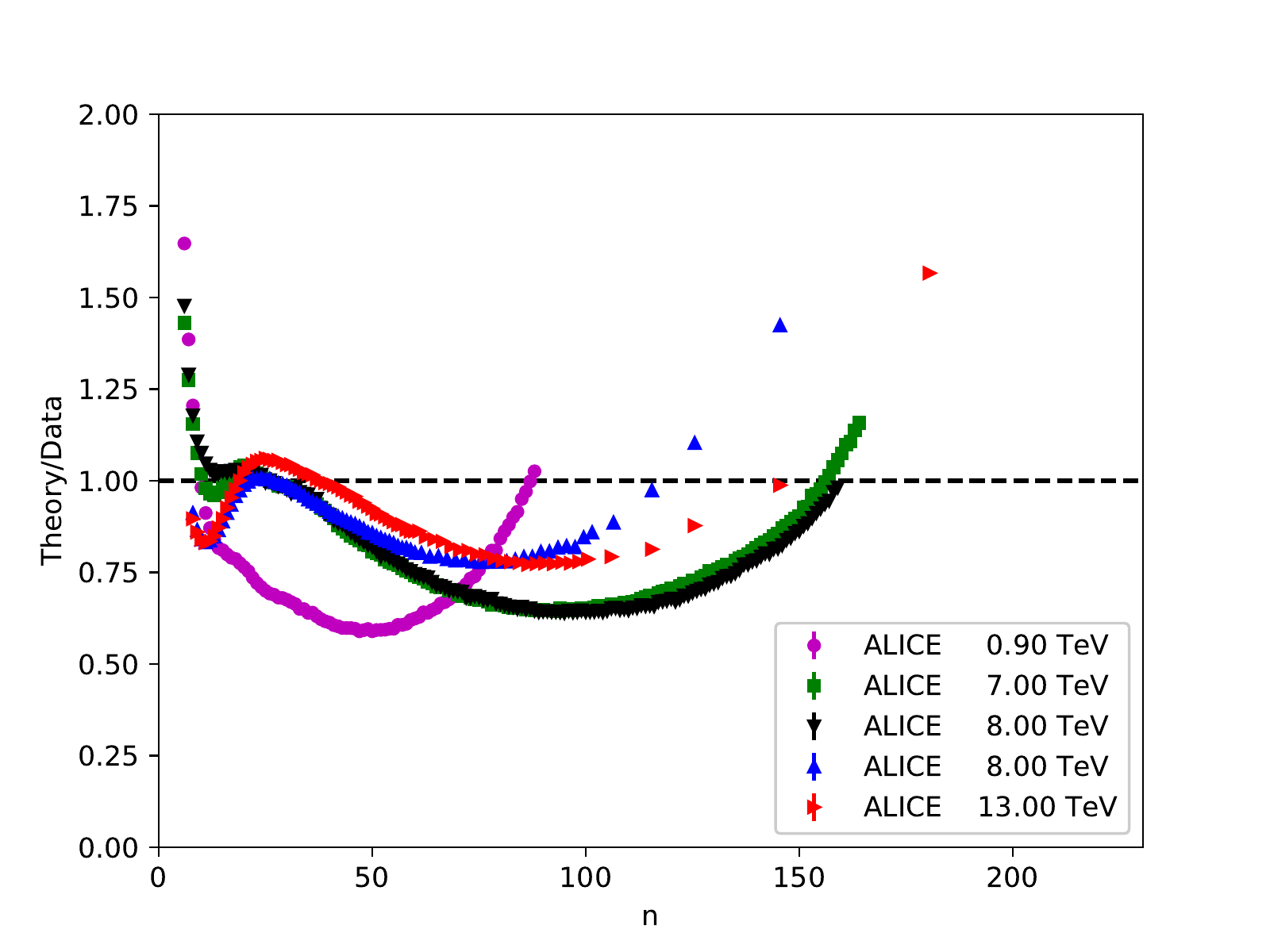}}\\
  \end{tabular}
%  {\includegraphics[width=.6\linewidth]{mult_100.png}}
\caption{Multiplicity distributions compared with data set II. 
  Data for $ \sqrt{s} = $  900,  7000 and 8000 GeV from ALICE (2017)
  \cite{alice17}. Data from ATLAS (2016) for 8000 \cite{atlas16a} and 13000 GeV
  \cite{atlas16b}.
  Solid lines: $P_{KL}(n)$.
  Dashed lines:	$P_{GL}(n)$.
  Left panel: $P(n)$ for different energies.
  Sets of points were multiplied by powers of ten for clarity. Right panel:
  Ratio Theory ($P_{KL}(n)$)/Data.}
\label{pn100}
\end{figure}

\begin{figure}
\begin{tabular}{cc}
  {\includegraphics[width=.5\linewidth]{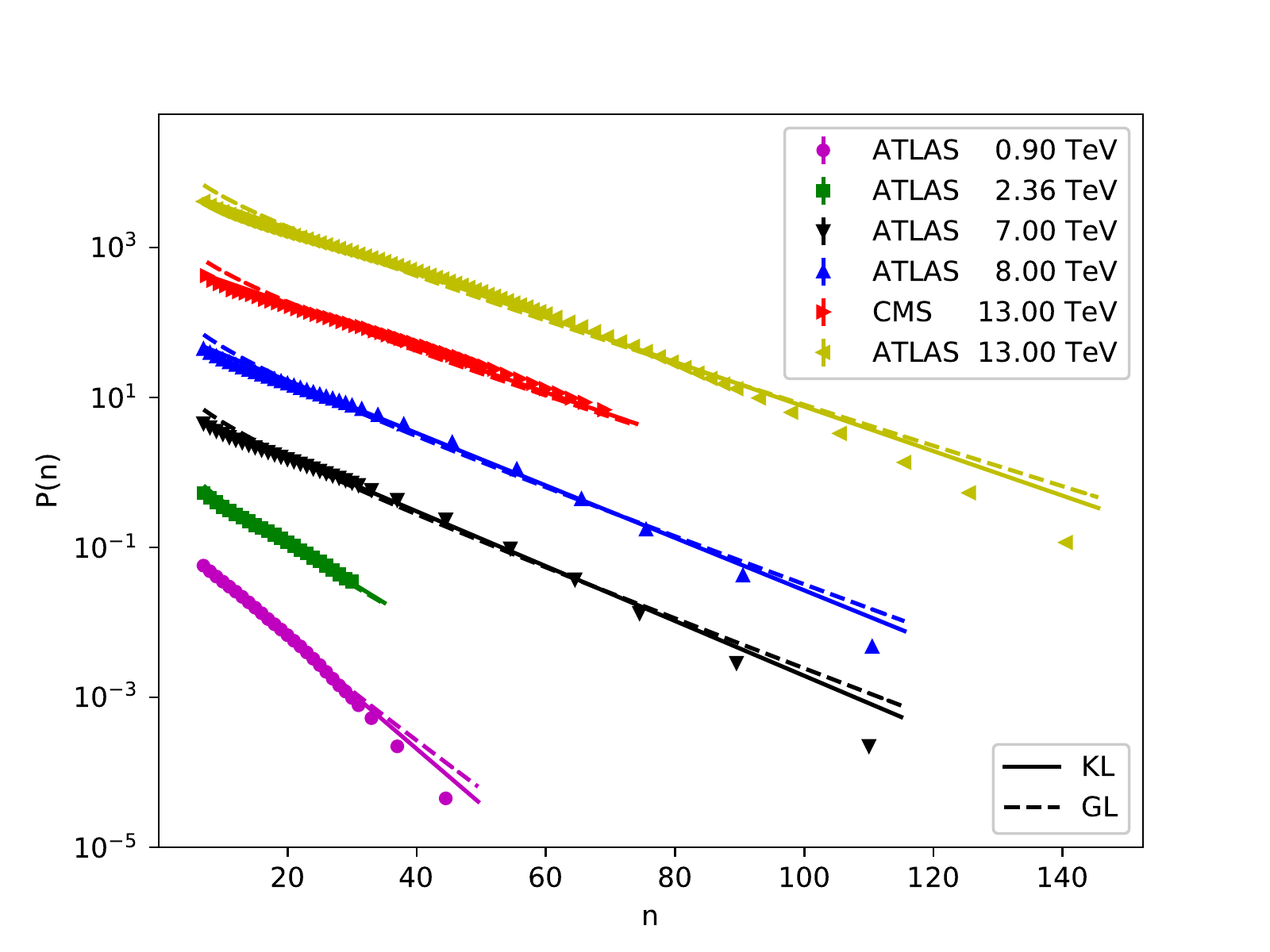}} &
  {\includegraphics[width=.5\linewidth]{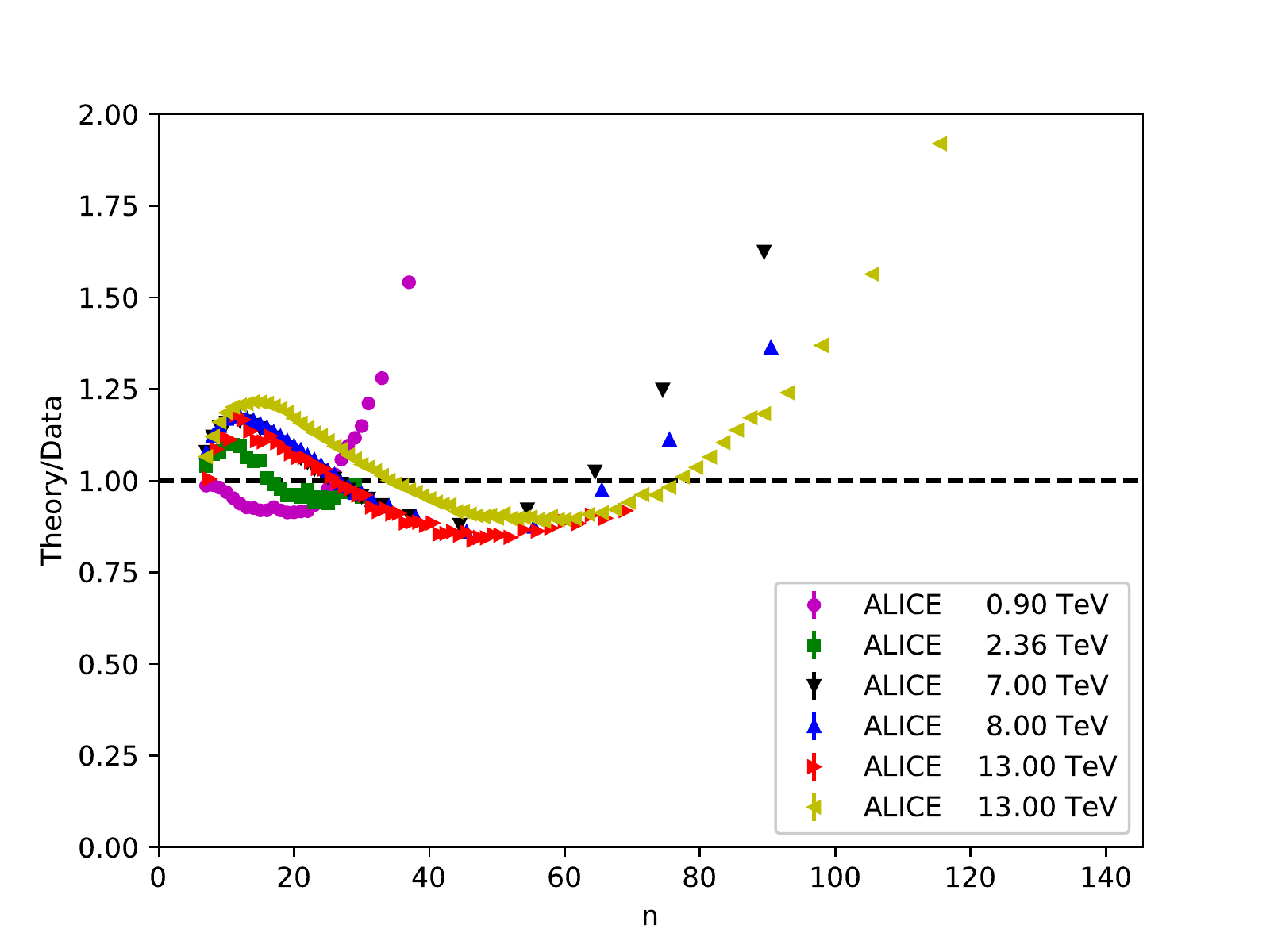}}\\
  \end{tabular}
%  {\includegraphics[width=.6\linewidth]{mult_500.png}}
\caption{Multiplicity distributions compared with data set III.
  Data for energies $\sqrt{s} = $ 900, 2360 and 7000 GeV from
  ATLAS (2011) \cite{atlas11}, for 8000 GeV  from ATLAS (2016) \cite{atlas16a}. 
Data for 13000 GeV from CMS (2018) \cite{cms18}
and from ATLAS (2016)  \cite{atlas16c}. Solid lines: $P_{KL}(n)$.
  Dashed lines:	$P_{GL}(n)$.
Left panel: $P(n)$ for different energies.
  Sets of points were multiplied by powers of ten for clarity. Right panel:
  Ratio Theory ($P_{KL}(n)$)/Data.}
\label{pn500}
\end{figure}

In Figs. \ref{pnmeio}, \ref{pn100} and \ref{pn500} we show the 
the fits of (\ref{klpn}) (solid lines) and (\ref{glpn}) (dashed lines)
to the data on multiplicity
distributions from the LHC.  For each figure we adjusted three numbers
$\Delta$, $q_0$ and $q_1$. They are listed in Table \ref{Table1}. 
As it can be seen both models overshoot the
data at large $n$ in all data sets. The model GL, as expected, overshoots
the data at small $n$. In Figs. \ref{pnmeio} and \ref{pn500} the theoretical
curves have essentially the correct shape. In contrast, in Fig. \ref{pn100},
in some cases, the data show a curvature which is absent in the theoretical
curves.  A careful comparison between set I and set II was done in
\cite{cms11}. The conclusion then was that set I is compatible with KNO
scaling while it is violated in set II. There seemed to be something different
happening in set II. Now set II has been enlarged with new data points from 
higher energies and the agreement with models KL and GL is better. From the
perspective of low x physics,  set III is the most interesting data sample
because the cut $p_{T} > 500 $ MeV removes a (significant ?) part of the
non-perturbative events. These events are not related to parton branching
or to the evolution equations. As can be seen in Table \ref{Table1}, the
value of $\Delta$ is larger, indicating a stronger dependence on the
energy $\sqrt{s}$, typical of perturbative physics. Finally, the overall
agreement between theory and data is better.

In \cite{gole} and \cite{gole2}  $P_{GL}(n)$ and $P_{KL}(n)$ were compared
with some experimental data but here we perform a systematic comparison to all
relevant available data. In those works the final goal was to compute the
entanglement entropy and the multiplicity distributions were a crucial      
ingredient.  To this end it is important to have an accurate description of
data.

The model presented here can be  improved in many ways. The multiplicity 
distributions are affected by fluctuations in several quantitites. The most
obvious one is the impact parameter $b$ \cite{raju11,beg}. Another one is
the fraction
(called inelasticity $K$) of the c.m.s energy which is converted into secondary
particles. It is typically $0.5$, but it changes from event to event and it
depends on the energy  according to a distribution 
$\chi[K(b,s)]$ \cite{beg,nuww}. The inclusion of these fluctuations will
certainly modify the shape of $P(n)$. Another aspect to
be considered is the number of sources which emit particles.  When we use
$P_{GL}(n)$ or $P_{KL}(n)$ above it is understood that there is only one emitting
system. However, in high energy proton-proton collisions there is a 
separation between particle production at low rapidities (mostly from the gluon
cloud of the projectiles) and at larger rapidities (mostly from the valence quarks).
So it is conceivable that the final, observed, $P(n)$,  is a sum of partial
multiplicity distributions. 

To summarize: multiplicity distributions deserve more attention from theorists.
In Refs. \cite{kale,gole,gole2}  simple distributions were derived from the
QCD evolution
equations. Here we have performed a comprehensive comparison between these
multiplicity distributions and LHC data. The results are encouraging but there is
still a lot to be done. 

\begin{table}
\begin{center}
%\begin{tabular}{|c|c|c|c|} 
\begin{tabular}{cccc}
\hline
\hline
Data Set & $\Delta$ & $q_{0}$ (GeV) & $q_{1}$ (GeV) \\  
\hline 
%\hline
I & 0.13 & 4.46 & 4.70 \\ 
\hline 
II & 0.13 & 0.02 & 0.01 \\ 
\hline 
III & 0.16 & 2.85 & 1.93  \\ 
\hline
\hline
\end{tabular} 
\caption{Fit parameters.}
\label{Table1}
\end{center}
\end{table}

%\begin{figure}
%\begin{tabular}{cc}
%{\includegraphics[width=.50\linewidth]{kabsgp.eps}} &
%{\includegraphics[width=.40\linewidth]{k-100.eps}}\\
%\end{tabular}
%\caption{Absorptive $\kappa$-factor for neutron production in real and
%  virtual photon-proton collisions. a) Results taken from  Ref.~\cite{pirner}.
%b) Results obtained with Eqs.(\ref{desdzgamma}), (\ref{f3}) and (\ref{f4}).}
%\label{kappa}
%\end{figure}

\begin{acknowledgments}

  The authors are grateful to G. Wilk for useful comments.  
  This work was  partially financed by the Brazilian funding
  agencies CNPq and  CAPES.

\end{acknowledgments}

\hspace{1.0cm}


\begin{thebibliography}{99}

  
\bibitem{fiete}     J.~F.~Grosse-Oetringhaus and K.~Reygers, 
                    J.\ Phys.\ G {\bf 37}, 083001 (2010). 
                    %[arXiv:0912.0023 [hep-ex]].

                    
\bibitem{atlas11}   G.~Aad \textit{et al.} [ATLAS],
                    %``Charged-particle multiplicities in pp interactions      
                    %measured with the ATLAS detector at the LHC,''              
                    New J. Phys. \textbf{13}, 053033 (2011).
                    %[arXiv:1012.5104 [hep-ex]].    
  
\bibitem{cms11}    V.~Khachatryan \textit{et al.} [CMS],
                   %``Charged Particle Multiplicities in $pp$                    
                   %Interactions at $\sqrt{s}=0.9$, 2.36, and 7 TeV,''             
                   JHEP \textbf{1101}, 079 (2011).
                   %[arXiv:1011.5531 [hep-ex]].        

\bibitem{atlas16a} G. Aad et al. [ATLAS Collaboration],
                   Eur. Phys. J. C {\bf 76}, 403 (2016).
                   % 8 TeV
                   
\bibitem{atlas16b} G. Aad et al. [ATLAS Collaboration],
                   Eur. Phys. J. C {\bf 76}, 502 (2016).
                   % 13 TeV
                   
\bibitem{atlas16c} G. Aad et al. [ATLAS Collaboration],
                   Phys. Lett. B {\bf 758}, 67 (2016).                   
                   % 13 TeV
                   
\bibitem{alice17}  S. Acharya {\it et al.} [ALICE Collaboration],
                   Eur. Phys. J. C {\bf 77},
                   852 (2017). % arXiv:1708.01435 [hep-ex].                      

\bibitem{cms18}    A.~M.~Sirunyan {\it et al.} [CMS Collaboration],
                   Eur.\ Phys.\ J.\ C {\bf 78}, 697 (2018).
                   %[arXiv:1806.11245 [hep-ex]].                                 

\bibitem{ugo98}    A.~Giovannini and R.~Ugoccioni,
                   Phys.\ Rev.\ D {\bf 59}, 094020 (1999);
                   Erratum: [Phys.\ Rev.\ D {\bf 69}, 059903 (2004)].

\bibitem{zbo}      I. Zborovsky, Eur. Phys. J. C {\bf 78}, 816 (2018).

\bibitem{bya}      M.~Biyajima and T.~Mizoguchi,
                   %``Unified description of multiplicity distributions            
                   %  based on the three-negative binomial distribution,''         
                   Int.\ J.\ Mod.\ Phys.\ A {\bf 34}, 1950203 (2019).
                   %[arXiv:1907.01967 [hep-ph]].   
  
\bibitem{kole}     Yuri V. Kovchegov and Eugene Levin,          
                   “ Quantum Chromodynamics at High Energies",
                   Cambridge Monographs on Particle Physics, Nuclear Physics
                   and Cosmology, Cambridge University Press, 2012 .

\bibitem{gel}      F. Gelis, E. Iancu, J. Jalilian-Marian and R. Venugopalan,
                   Ann. Rev. Nucl. Part. Sci. {\bf 60}, 463 (2010). 
                                     
\bibitem{schenke}  B.~Schenke, P.~Tribedy and R.~Venugopalan,
                   Phys.\ Rev.\ C {\bf 89}, 024901 (2014).
                   %[arXiv:1311.3636 [hep-ph]].                                  

\bibitem{nara}     A.~Dumitru, D.~E.~Kharzeev, E.~M.~Levin and Y.~Nara,
                   %``Gluon Saturation in $pA$ Collisions at the LHC: KLN Model
                   %Predictions For Hadron Multiplicities,''                     
                   Phys.\ Rev.\ C {\bf 85}, 044920 (2012).
                   %[arXiv:1111.3031 [hep-ph]].                        
%\bibitem{lappi}     T.~Lappi,
                     %``Energy dependence of  saturation scale and the charged  
%                    %multiplicity in pp and AA collisions,''                    
%                    Eur.\ Phys.\ J.\ C {\bf 71}, 1699 (2011).                   
%                    %[arXiv:1104.3725 [hep-ph]].      

\bibitem{raju11}   P.~Tribedy and R.~Venugopalan,
                   %``Saturation models of HERA data and hadron distributions    
                   Nucl.\ Phys.\ A {\bf 850}, 136 (2011).
                   Erratum: [Nucl.\ Phys.\ A {\bf 859}, 185 (2011)].
                   %[arXiv:1011.1895 [hep-ph]].
                   
\bibitem{glitter}  F.~Gelis, T.~Lappi and L.~McLerran,
                   Nucl.\ Phys.\ A {\bf 828}, 149 (2009).
                   %[arXiv:0905.3234 [hep-ph]].   
                   

\bibitem{mumu}     T.~Liou, A.~H.~Mueller and S.~Munier,
                   %``Fluctuations of the multiplicity of produced particles     
                   %in onium-nucleus collisions,''                               
                   Phys.\ Rev.\ D {\bf 95}, 014001 (2017).
                   %[arXiv:1608.00852 [hep-ph]].                                 

\bibitem{giaca}    L.~Dominé, G.~Giacalone, C.~Lorcé, S.~Munier and S.~Pekar,
                   %``Gluon density fluctuations in dilute hadrons,''            
                   Phys.\ Rev.\ D {\bf 98}, 114032 (2018).
                   %[arXiv:1810.05049 [hep-ph]].
                   
\bibitem{dumi}     A.~Dumitru and V.~Skokov,
                   %``Fluctuations of the gluon distribution..."                 
                   Phys.\ Rev.\ D {\bf 96}, 056029 (2017).
                   %[arXiv:1704.05917 [hep-ph]].                                 

\bibitem{shab}     G.~H.~Arakelyan, Y.~M.~Shabelski and A.~G.~Shuvaev,
                   %``Multiplicity distribution of gluons in pQCD,''             
                   Eur.\ Phys.\ J.\ C {\bf 80}, 592 (2020).
                   %[arXiv:2003.03275 [hep-ph]].  

\bibitem{khoze}    V.~V.~Khoze, F.~Krauss and M.~Schott,
                   JHEP {\bf 2004}, 201 (2020);
                   %[arXiv:1911.09726 [hep-ph]].                    
                   V.~V.~Khoze, D.~L.~Milne and M.~Spannowsky,
                   %``Searching for QCD Instantons at Hadron Colliders,''       
                   arXiv:2010.02287 [hep-ph].                   

\bibitem{beg}      P.~C.~Beggio and F.~R.~Coriolano,
                   %``Energy dependence of the inelasticity in                   
                   %$pp/p{\overline{p}}$ collisions from experimental information
                   %on charged particle multiplicity distributions,''           
                   Eur.\ Phys.\ J.\ C {\bf 80},  437 (2020).
                   %[arXiv:2004.06839 [hep-ph]].
                   
\bibitem{agga}     R.~Aggarwal and M.~Kaur,
                   %``Compelling evidence of oscillatory behaviour of            
                   %hadronic multiplicities in the shifted Gompertz distribution"
                   Adv.\ High Energy Phys.\  {\bf 2020}, 5464682 (2020).
                   %[arXiv:1911.03896 [hep-ph]].    
                   
\bibitem{sha}      S.~Sharma and M.~Kaur,
                   Phys.\ Rev.\ D {\bf 99}, 096016 (2019).
                   %[arXiv:1903.10761 [hep-ph]].

\bibitem{gre18}    M.~Rybczynski, G.~Wilk and Z.~Włodarczyk,
                   %``Intriguing properties of multiplicity distributions,''
                   Phys.\ Rev.\ D {\bf 99}, 094045 (2019).
                   
\bibitem{gre19}    M.~Rybczyński, G.~Wilk and Z.~Włodarczyk,
                   %``A look at multiplicity distributions''                  
                   Ukr.\ J.\ Phys.\  {\bf 64}, 738 (2019).
                   %[arXiv:1906.11531 [hep-ph]].   

\bibitem{gre20}     H.~W.~Ang, A.~H.~Chan, M.~Ghaffar, M.~Rybczyński,
                    G.~Wilk and Z.~Włodarczyk,
                    %``A look at multiparticle production via modified      
                    %combinants,''                                             
                    Eur.\ Phys.\ J.\ A {\bf 56}, 117 (2020).
                    %[arXiv:1908.11062 [hep-ph]].
                    
\bibitem{adrian}   A.~Dumitru and E.~Petreska,
                   %``KNO scaling from a nearly Gaussian action                  
                   %for small-x gluons,''                                        
                   arXiv:1209.4105.

\bibitem{duna}     A.~Dumitru and Y.~Nara,
                   %``KNO scaling of fluctuations in pp..."                      
                   Phys.\ Rev.\ C {\bf 85}, 034907 (2012).
                   %[arXiv:1201.6382 [nucl-th]].                     

\bibitem{lund}     C.~Bierlich, S.~Chakraborty, G.~Gustafson and L.~Lönnblad,
                   %``Setting the string shoving picture in a new frame,''
                   arXiv:2010.07595 [hep-ph].

\bibitem{lubi}     E.~Levin and M.~Lublinsky,
                   Nucl.\ Phys.\ A {\bf 730}, 191 (2004).
                   %[hep-ph/0308279].
                   
\bibitem{kale}     D.~E.~Kharzeev and E.~M.~Levin,
                   Phys.\ Rev.\ D {\bf 95}, 114008 (2017).
                   %[arXiv:1702.03489 [hep-ph]].   

\bibitem{gole}     E. Gotsman and E. Levin,
                   Phys.\ Rev.\ D {\bf 102}, 074008 (2020).
                   %[arXiv:2006.11793 [hep-ph]];
                   
\bibitem{gole2}    E. Gotsman and E. Levin,
                   arXiv:2008.10911 [hep-ph].


\bibitem{lelu}     J. Bartels, E. Gotsman, E. Levin, M. Lublinsky and U. Maor,
                   Phys. Lett. B {\bf 556}, 114 (2003).

\bibitem{babi}     F.~Carvalho, F.~O.~Duraes, V.~P.~Goncalves and
                   F.~S.~Navarra,
                   Mod.\ Phys.\ Lett.\ A {\bf 23}, 2847 (2008).
                   %[arXiv:0705.1842 [hep-ph]].     
                   
\bibitem{kn}       D.~Kharzeev and M.~Nardi,
                   %``Hadron production in nuclear collisions...                 
                   Phys.\ Lett.\ B {\bf 507}, 121 (2001).
                   %  [nucl-th/0012025].                                    

\bibitem{kln}      D.~Kharzeev, E.~Levin and M.~Nardi,
                   %``Color glass condensate at the LHC:...
       	           Nucl.\ Phys.\ A {\bf 747}, 609 (2005).
                   %[hep-ph/0408050].    


\bibitem{dosch}    H.~G.~Dosch, E.~Ferreira and A.~Kramer,
                   %``Nonperturbative QCD treatment of high-energy
                   %hadron hadron scattering,''
                   Phys.\ Rev.\ D {\bf 50}, 1992 (1994). 
                   %[hep-ph/9405237].

\bibitem{roy}      R.~A.~Lacey, P.~Liu, N.~Magdy, M.~Csanád, B.~Schweid,
                   N.~N.~Ajitanand, J.~Alexander and R.~Pak,
                   Universe {\bf 4}, 22 (2018).
                   % [arXiv:1601.06001 [nucl-ex]].  

\bibitem{nuww}     F.~S.~Navarra, O.~V.~Utyuzh, G.~Wilk and Z.~Wlodarczyk,
                   Phys.\ Rev.\ D {\bf 67}, 114002 (2003); 
                   %[hep-ph/0301258].  
                   G.~N.~Fowler et al., 
                   Phys.\ Rev.\ C {\bf 40}, 1219 (1989).
                   
%\bibitem{tri}      D. N. Triantafyllopoulos,
%                   Nucl. Phys. B {\bf 648}, 293 (2003).
 

%\bibitem{prasa}    M.~Praszalowicz,
%                   Phys.\ Lett.\ B {\bf 704}, 566 (2011).
%                   %[arXiv:1101.6012 [hep-ph]].

                   
%\bibitem{india}    P.~Sarma and B.~Bhattacharjee,
%                   Phys.\ Rev.\ C {\bf 99}, 034901 (2019).  
%                   %[arXiv:1902.09124 [hep-ph]].

%\bibitem{kovner}   A. Kovner, E. Levin and M. Lublinsky, arXiv:1605.03251.
 
%\bibitem{polones}  M. Praszalowicz, Phys. Lett. B {\bf 704}, 566 (2011). 
%                   %arXiv:1101.6012 [hep-ph]]

%\bibitem{nuww}     F.~S.~Navarra, O.~V.~Utyuzh, G.~Wilk and Z.~Wlodarczyk,   
%                   Phys.\ Rev.\ D {\bf 67}, 114002 (2003). 
%                   %[hep-ph/0301258].

%\bibitem{lappi}     T.~Lappi,
%                    %``Energy dependence of the saturation scale and the charged
%                    %multiplicity in pp and AA collisions,''
%                    Eur.\ Phys.\ J.\ C {\bf 71}, 1699 (2011).  
%                    %[arXiv:1104.3725 [hep-ph]].                 

%\bibitem{drem}     I. M. Dremin and V. A. Nechitailo,
%                   Phys. Rev. D {\bf 70},  034005 (2004).

  
\end{thebibliography}
\end{document}